\documentclass[a4paper,11pt]{article}
\usepackage{jcappub}

\usepackage{mathrsfs}
\usepackage{latexsym}
\usepackage{amsmath}
\usepackage{amssymb}
\usepackage{graphicx}
\usepackage{subfigure}
\usepackage{dcolumn}
\usepackage{bm}
\usepackage[normalem]{ulem}
\usepackage{color}
\usepackage{comment}
\usepackage{booktabs}

\arxivnumber{2509.07831}
\title{\boldmath Is GW190521 a gravitational wave echo of wormhole remnant from another universe?}

\author[a,d,e]{Qi Lai,}
\author[b]{Qing-Yu Lan,}
\author[b]{Hao-Yang Liu,}
\author[a,b]{Yu-Tong Wang,}
\author[a,b,c,d]{Yun-Song Piao}

\affiliation[a]{School of Fundamental Physics and Mathematical Sciences, Hangzhou Institute for Advanced Study, UCAS, Hangzhou 310024, China}
\affiliation[b]{School of Physical Sciences, University of Chinese Academy of Sciences, Beijing 100049, China}
\affiliation[c]{International Center for Theoretical Physics Asia-Pacific, Beijing/Hangzhou, China}
\affiliation[d]{Institute of Theoretical Physics, Chinese Academy of Sciences, P.O. Box 2735, Beijing 100190, China}
\affiliation[e]{University of Chinese Academy of Sciences, Beijing 100190, China}

\emailAdd{laiqi23@mails.ucas.ac.cn}
\emailAdd{yspiao@ucas.ac.cn}

\abstract{
A particularly compelling aspect of the GW190521 event detected by the LIGO--Virgo--KAGRA (LVK) collaboration is that it has an extremely short duration, and lacks a clearly identifiable inspiral phase usually observed in the binary black holes (BBHs) coalescence. In this work, we hypothesize that GW190521 might represent a single, isolated gravitational wave (GW) echo pulse from the wormhole, which is the postmerger remnant of BBHs in another universe and connected to our universe through a throat. The ringdown signal after BBHs merged in another universe can pass through the throat of wormhole and be detected in our universe as a short-duration echo pulse. Our analysis results indicate that our model yields a network signal-to-noise ratio comparable to that of the standard BBHs merger model reported by the LVK collaboration. For GW190521, Bayesian model selection yields $\ln \mathcal{B}^{\text{Echo}}_{\text{BBH}} \simeq -2.9$, indicating that the data favor the BBH hypothesis over our echo-for-wormhole model.
}


\begin{document}
\maketitle
\flushbottom


\section{Introduction}

As of now, the LIGO--Virgo--KAGRA (LVK) collaboration has reported 218 gravitational wave (GW) events~\cite{LIGOScientific:2018mvr,LIGOScientific:2020ibl,LIGOScientific:2021usb,KAGRA:2021vkt,LIGOScientific:2025slb}. Their successful detection has opened new avenues for comprehending our universe, particularly physics in strong gravitational fields.

To date, GW events have been confidently attributed to the mergers of binary black holes (BBHs) or binary neutron stars (BNSs), with their signals displaying the well-understood inspiral-merger-ringdown (IMR) morphology~\cite{LIGOScientific:2018mvr,LIGOScientific:2020ibl,LIGOScientific:2021usb,KAGRA:2021vkt,LIGOScientific:2025slb}. However, among the cataloged events, GW190521 is an unusual and intriguing case that
might challenge this standard
paradigm~\cite{LIGOScientific:2020iuh}, since it has an extremely
short duration (the order of 0.1 seconds) and lacks a clearly
identifiable inspiral phase usually observed in the mergers of
BBHs. This lack raises questions about its physical origin~\cite{KAGRA:2021duu}. Though
the LVK collaboration has interpreted GW190521 as the
merger of BBHs with masses in $85^{+21}_{-14}M_{\odot}$ and
$66^{+17}_{-18}M_{\odot}$, resulting in a postmerger remnant of
$142^{+28}_{-16}M_{\odot}$~\cite{LIGOScientific:2020iuh},
see also relevant
e.g.\ Refs.~\cite{Fishbach:2020qag,Nitz:2020mga,Estelles:2021jnz,Capano:2022zqm},
it is in tension with established astrophysical models of stellar
evolution. Thus some alternative interpretations such as primordial black holes, cosmic strings, new light particles, and even horizonless compact objects have been
presented~\cite{Clesse:2020ghq,Aurrekoetxea:2023vtp,CalderonBustillo:2020fyi,Sakstein:2020axg,Abedi:2021tti,CalderonBustillo:2022cja}.

In parallel, to investigate possible macroscopic signatures of
quantum gravity and the longstanding black hole information
paradox, a variety of horizonless exotic compact objects have been
investigated, e.g.\cite{Cardoso:2019rvt} for a recent review. The
wormhole represents such an object connecting either two separate
universes or two distant regions in a single universe through a
throat~\cite{Morris:1988cz,Morris:1988tu}, see also
\cite{Poisson:1995sv,Hochberg:1998ha,Armendariz-Picon:2002gjc,Garcia:2011aa,Kanti:2011jz,Damour:2007ap}.
It has been shown that if the remnant object after BBHs merged is
a wormhole, the corresponding ringdown burst propagating toward
the other side of wormhole can be reflected back by the photon
sphere on the far end, yielding a sequence of delayed pulses,
known as echoes~\cite{Cardoso:2016rao}. The properties of echoes,
such as their amplitude and time delay, carry information about
the underlying physical characteristics of corresponding
wormholes,
e.g.\cite{Cardoso:2017cqb,Bueno:2017hyj,Wang:2018mlp,Correia:2018apm,GalvezGhersi:2019lag,Li:2019kwa,Bronnikov:2019sbx,Liu:2020qia,Yang:2024prm}
; see also Refs.~\cite{Siemonsen:2024snb,Ma:2022xmp,Annulli:2021ccn,Wang:2019rcf,Maggio:2019zyv}.

A particularly compelling aspect of GW190521 is the lack of
precursor signal prior to the main burst, which hints the
possibility that it might not originate from a standard
coalescence process of BBHs. In this work, we hypothesize that
GW190521 might represent a single, isolated GW echo pulse from the
wormhole, which is the postmerger remnant of BBHs in another
universe and connected to our universe through a throat. The
postmerger ringdown signal passes through the throat of wormhole
and penetrate the photon sphere barrier in the side of our
universe, and could then be detected in our universe as a
short-duration burst lacking a pre-merger phase. To test this
hypothesis, we compute signal-to-noise ratio (SNR) of our model,
and compare it with the standard BBHs model using the Bayesian
analysis. Our results suggest that the BBHs hypothesis is still
preferred, but our echo-for-wormhole hypothesis can be a viable
alternative explanation for GW190521.

\section{Model}\label{Sec:Model and Waveform}

\subsection{Our echo-for-wormhole model}\label{Subsec:model}

\begin{figure*}[tbp]
  \centering
  \includegraphics[width=0.9\textwidth]{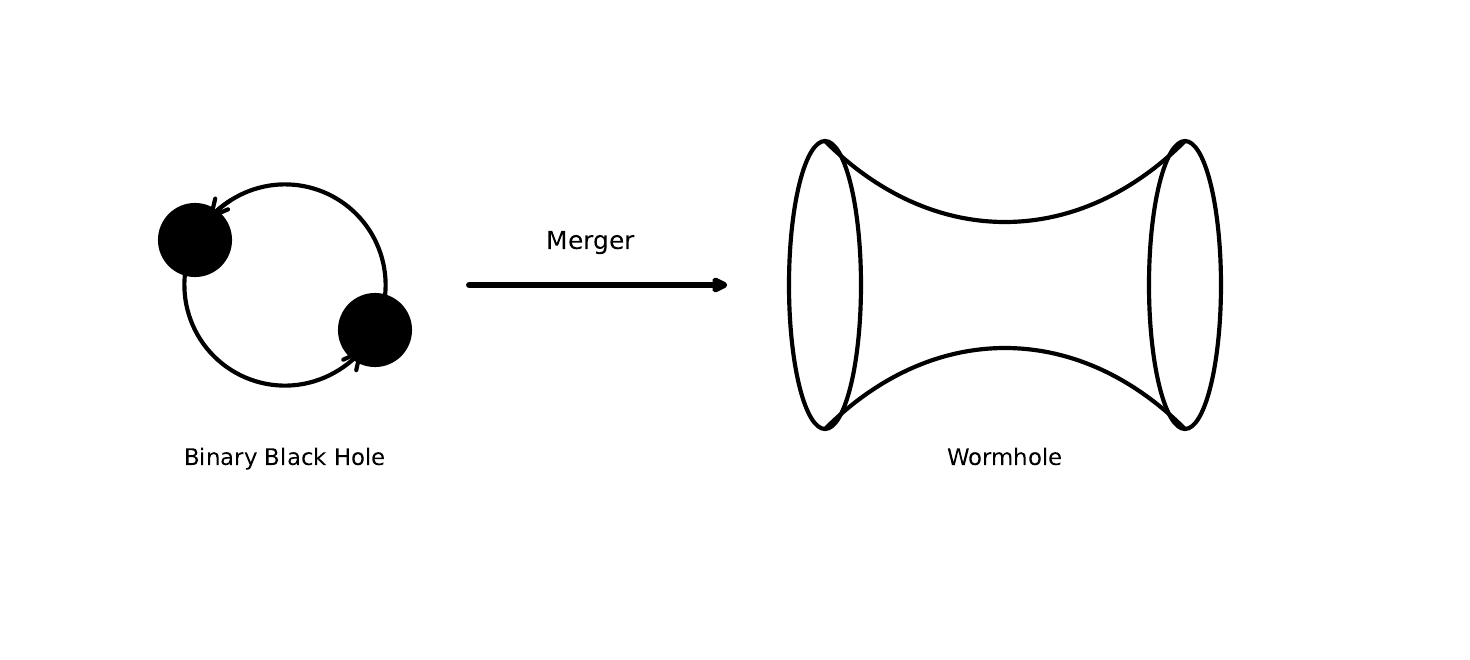}\\[1ex]
  \includegraphics[width=0.95\textwidth]{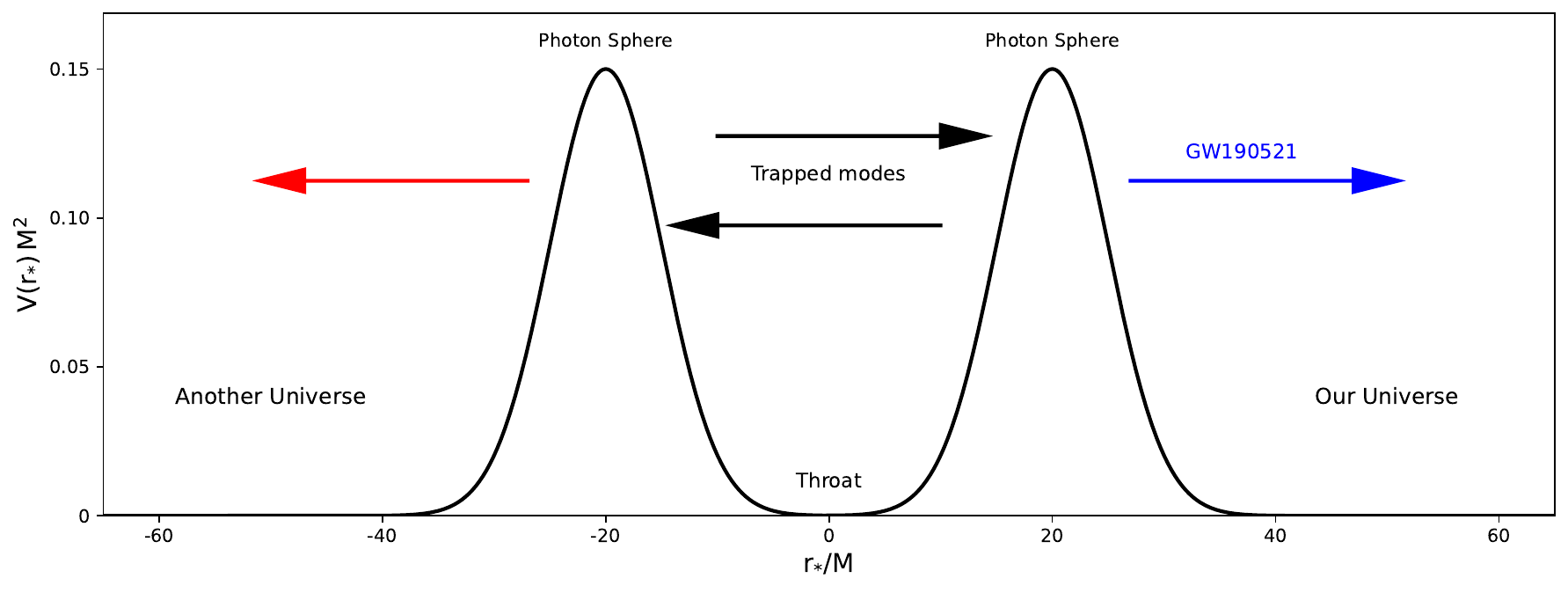}
\caption{Schematic of the echo interpretation of GW190521. Top: A
binary black hole merger forms a wormhole. Bottom: The black
mirror barriers represent a wormhole connecting two universes. A
portion of the merger-generated ringdown GWs enters the wormhole.
These GW modes operates between mirror barriers due to the
reflections of barriers, and certain modes tunnel through the
right-side barrier into our universe, leading to a series of
echoes in our universe. The first echo pulse (blue arrow) was
observed as GW190521.}
\label{wormhole_double_barrier}
\end{figure*}

As a proof-of-principle model, we consider a Schwarzschild--like
Morris--Thorne wormhole \cite{Morris:1988cz,Morris:1988tu}, see also recent~\cite{Cardoso:2016rao}, in which the
Schwarzschild metrics of both sides are glued at $r=r_{0}>r_{S}$,
and $r_{0}$ is the radius of wormhole throat and $r_{S}$ is the
Schwarzschild radius. We note that, the remnant of GW190521 is highly spinning with $\chi_f = 0.72^{+0.09}_{-0.12}$~\cite{LIGOScientific:2020iuh}. Here, we adopt a nonspinning setup for simplicity and thus do not model spin-induced modifications to the effective potential or the echo spectrum, which constitutes a limitation of our model. As shown exactly in
Ref.\cite{Bueno:2017hyj}, the relevant issue on post-ringdown GW
signals can be regarded equivalently as a one-dimensional
scattering problem with a frequency-domain wave equation: 
\begin{equation}
    \frac{d^2{\tilde\Psi}}{d{r_{*}}^{2}}+\left(\omega^2+V(r_{*},\omega)\right){\tilde\Psi}=0
\end{equation}
where the boundary condition is
$\lim_{r_{*}\longrightarrow\pm\infty}{\tilde\Psi}(r_{*})\sim
e^{\pm i\omega r_{*}}$ and the effective potential is a pair of
mirror barriers (Schwarzschild angular momentum barrier) with
their peaks representing photon spheres. The time-domain solution
is $\Psi(t)={\frac{1}{2\pi}}\int d\omega\, {\tilde\Psi}(\omega)$.
The central region between the mirror barriers represents the
throat of wormhole, while the regions beyond the barriers
correspond to our universe and another universe, respectively, as
illustrated in Fig.\ref{wormhole_double_barrier}.

In our model, see Fig.\ref{wormhole_double_barrier}, the BBHs inspiraled and merged in the left-side universe, and their postmerger object is a Schwarzschild-like wormhole connecting it to our universe. As a result, the postmerger GW ringdown signal, which is dominated by the frequency content of light-ring perturbations, will not only propagate outward at the speed of light, but also enter the wormhole throat and penetrate the right-side barrier into our universe. It can be expected that the reflections of such a pair of mirror barriers for the ringdown pulse will yield a series of echo pulses with sequentially suppressed amplitudes, and the frequency of each echo decreases until it reaches the wormhole
quasinormal modes (QNMs)~\cite{Cardoso:2019rvt}. Therefore, the echo signal can be described as a sum of wormhole QNMs at late times~\cite{Bueno:2017hyj}
\begin{equation}
    \Psi(t)\approx e^{-i\omega_0t}
\left(\sum_{n=-\infty}^{+\infty}c_ne^{-n\pi t/L}e^{{\rm
Im}(\omega_{n})t}\right),
\end{equation} 
where 
\begin{equation}
    L\simeq
2\int_{r_{0}}^{\frac{3r_S}{2}}\frac{dr}{1-r_S/r}
\end{equation} 
represents the separation between the mirror barriers, $\omega_n$ is the
QNF of wormhole, and the coefficients
$c_n$ can be written as 
\begin{equation}
    c_n={\frac{1}{2L}}\int_0^{2L}dt
\Psi_{1^{\rm st}{\rm echo}}e^{i\omega_n t}, 
\end{equation}
where
$\Psi_{1^{\rm st}{\rm echo}}$ is the waveform of the first echo.
The interval between two echoes is approximately $\Delta t_{\pm
echo}\simeq 2L$, however, if the postmerger object is a wormhole
collapsing into a black hole, the change of interval after a
period is \cite{Wang:2018mlp} 
\begin{equation}
    \Delta t_{\rm echo}\simeq {2L\over 1-{\Delta L\over L}}, 
\end{equation} 
increasing with time, where $\Delta L$ is the change of the separation between mirror
barriers.

The postmerger ringdown signal closely resembles that of a black
hole with Schwarzschild radius $r_{S}$. The waveform of the first
echo is primarily governed by the wormhole QNF that lies closest to the corresponding black hole QNF
$\omega_0^{\rm BH}$, thus it is natural to model the first echo as
a Gaussian-like wave packet 
\begin{equation}\label{1echo} 
    \Psi_{1^{\rm st}{\rm echo}}\sim e^{i\omega_0^{\rm BH}(t-t_*)}e^{-{(t-t_*)^2\over 2\beta^2}},
\end{equation}  
where $\beta$ characterizes the width of the
Gaussian-shaped echo pulse~\cite{Maselli:2017tfq}. This result has been also
verified numerically in Ref.~\cite{Bueno:2017hyj}.

\subsection{Template of waveform}\label{Subsec:waveform}

It can be postulated that, after the merger of binary black holes,
a postmerger wormhole forms as an intermediate state, which
rapidly pinches off (after the postmerger ringdown signal
penetrates the right-side barrier into our universe) and
eventually collapses into a black hole \cite{Wang:2018mlp}, or
that the amplitudes of subsequent echoes fall below the
sensitivity threshold of current detectors~\cite{Abedi:2021tti}.
In either case, only the first echo pulse can be detected in our
universe\footnote{Recently, numerous studies have focused on the
search for echoes after the BBHs inspiraled and merged in one
single universe,
e.g.\cite{Abedi:2016hgu,Westerweck:2017hus,Lo:2018sep,Nielsen:2018lkf,Wang:2019szm,Uchikata:2019frs,Wang:2020ayy,LIGOScientific:2020tif,Ren:2021xbe,Abedi:2021tti,Abedi:2022bph,Wu:2023wfv,Uchikata:2023zcu},
showing no evidence for echo, however, here what we consider is to
take the GW190521 event itself as an echo (detected in our
universe) of postmerger wormhole in another universe in which the
BBHs inspiraled and merged.}. According to (\ref{1echo}), we use a
simplified \texttt{sine-Gaussian} template\footnote{ The
\texttt{sine-Gaussian} template has been widely used in the search
for GW echoes, e.g.Refs.\cite{Tsang:2018uie,Tsang:2019zra} and
LIGO-Virgo-KAGRA collaboration's work with GWTC-3
\cite{LIGOScientific:2021sio}.}: 
\begin{eqnarray}
    &h(t)&\left(\sim e^{2\pi
if_{c}(t-t_c)}e^{-{(t-t_c)^2\over 2\beta^2}}\right)\nonumber\\
&=&\mathcal{A}\mathrm{cos}\left(2\pi f_{c}
(t-t_{c})+\varphi\right)e^{-\frac{(t-t_{c})^{2}}{2\beta^{2}}},
\label{Waveformecho}
\end{eqnarray} 
where $\varphi$ is the reference phase,
$f_{c}$ denotes the central frequency of echo,
and $t$ is defined relative to trigger time
(1242442967.459 GPS time) as shown in Fig.\ref{corner_echo}.

The amplitude $\mathcal{A}$ depends on the energy of the GWs
emitted by the BBHs merger in another universe. However, in our
case, the total energy emitted by the source is unknown, which
causes the amplitude parameter inferred from the Bayesian analysis
to encode both the intrinsic strength of the source and its
distance to the detector. To facilitate the parameter estimation,
we fix a reference strain amplitude $A_{ref}=10^{-19}$ at a
fiducial distance of 1Mpc, the physical amplitude $\mathcal{A}$ is
then rescaled by a dimensionless factor
$A_{m}=\mathcal{A}/A_{ref}$ and set as a free parameter in our
analysis. This normalization not only improves numerical stability
but also yields a consistent scale for comparing with
LVK BBHs models.

\section{Results}\label{Sec:Data analysis}

\subsection{Our echo-for-wormhole model}

\begin{table}[tbp]
\centering
\begin{tabular}{ccc}
\hline\hline
Parameters & Prior range  \\
\hline
$A_{m}$  & $[0.0001,0.17]$ \\
$\beta$   & $[0.0058,0.026]$ \\
$f_{c}$  & $[52,70]$ \\
$\psi$ & $[0,\pi]$  \\
$\varphi$ & $[0,2\pi]$  \\
$\iota$ & $[0,\pi]$  \\
Geocent time  &  $[1242442967.3, 1242442967.5]$  \\
\hline\hline
\end{tabular}
\caption{Prior ranges of parameters for our echo-for-wormhole
waveform (\ref{Waveformecho}). Here, $\psi$ is the polarization,
while $\iota$ is the inclination refers to the orientation of binary relative
to the line of sight. Geocent time parameter specifies the
arrival time of the echo pulse at the Earth. The physical
interpretations of other parameters are presented in
Sec.\ref{Subsec:waveform}} \label{Prior_echo}
\end{table}
\begin{figure*}[tbp]
\includegraphics[width=0.95\textwidth]{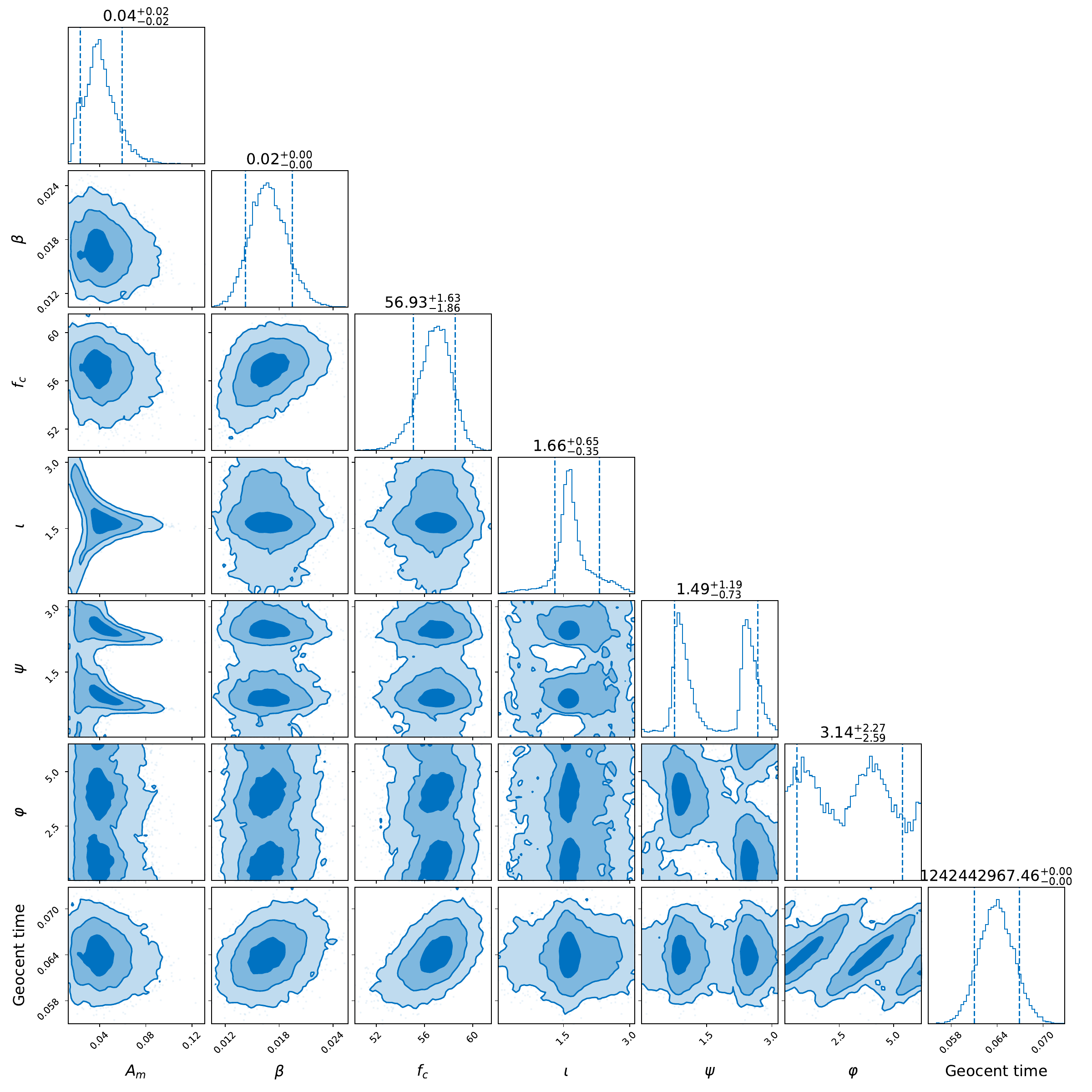}
\caption{The posterior distribution of parameters for our
echo-for-wormhole model. In this analysis, the sky location was
fixed to the best-fit values reported by ZFT19abanrhr. }
\label{corner_echo}
\end{figure*}

The priors of our echo-for-wormhole model are shown in
Tab.\ref{Prior_echo}. To avoid an implicit prior preference for large amplitudes, we adopt a log-uniform prior for $A_m$ within the specified range, while all other parameters are assigned uniform priors within their respective ranges.
In the initial parameter estimation, we find
that the maximum-posterior sky location inferred is consistent
with that obtained using the LVK BBHs model, and also with
the localization reported for ZTF19abanrhr, as shown in Appendix~\ref{App:Sky location}.
Thus we fix the location parameters with ZFT19abanrhr localization
(RA, DEC)$=(3.35847,~0.60781)$ to break the degeneracy between
arrival time and sky position. The marginalized posterior
probability distributions for all parameters in our model are
presented in Fig.~\ref{corner_echo}, where
$f_{c}=56.93^{+1.63}_{-1.86}$ Hz and $\beta\approx 0.02$. This result is consistent with the narrow-band feature of an echo pulse produced by a wormhole in the frequency domain.

\begin{figure*}[tbp]
\includegraphics[width=0.45\textwidth]{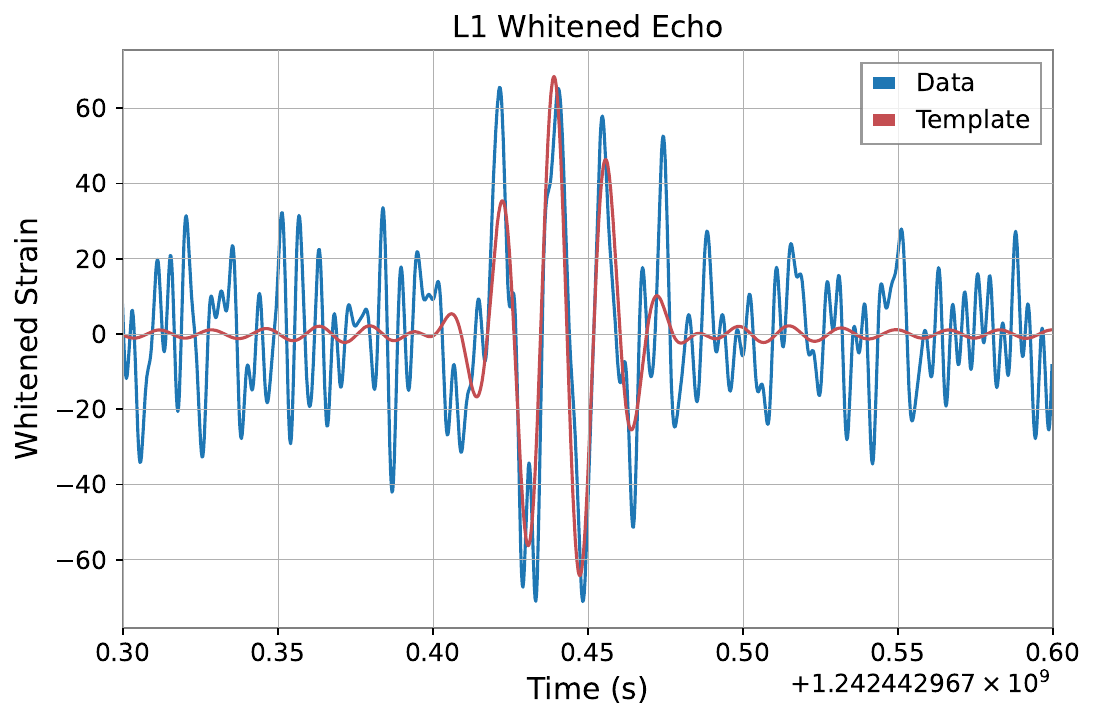}
\includegraphics[width=0.45\textwidth]{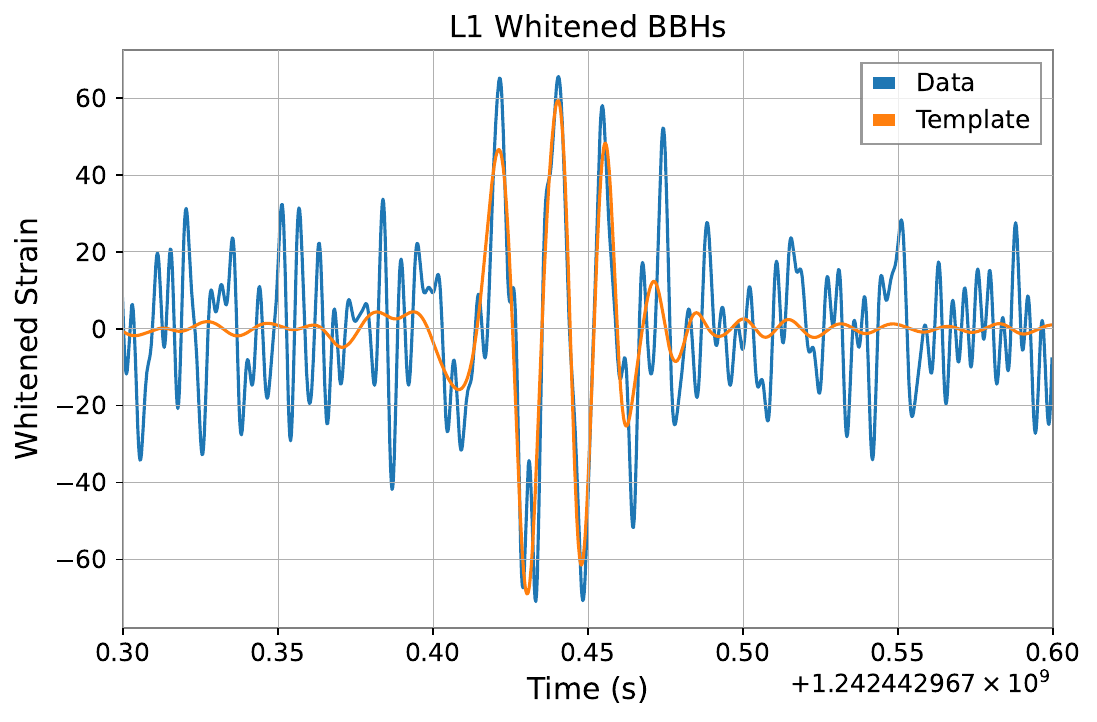}
\caption{The maximum-likelihood waveform reconstructions for the
LIGO Livingston (L1) detector in the time-domain with the GPS
time. The left panel shows the result for our echo-for-wormhole
model (Red), while the right panel corresponds to the BBHs model
(Orange) using the \texttt{IMRPhenomXPHM} waveform. }
\label{WaveformL1}
\end{figure*}

In Fig.\ref{WaveformL1}, we plot the best-fit echo-for-wormhole
waveform (\ref{Waveformecho}) with the strain data in L1 detector
in time-domain, and also the BBHs waveform for a comparison, where owing to the short duration of GW190521 (approximately $0.1$~s), we additionally bandpass the strain data between 20--256~Hz to
suppress high frequencies, resulting in an almost 5-cycle waveform. Here, to facilitate a direct comparison between the two models, we whiten both the strain data and the corresponding waveforms according to
\begin{equation}
    \tilde{h}_{w}(f)=\frac{\tilde{h}(f)}{\sqrt{\tilde{S}_{n}(f)}},\label{Eq:whiten}
\end{equation}
where $\tilde{h}_{w}(f)$ denotes the whitened waveform in the frequency domain; $\tilde{S}_{n}(f)$ is the one-sided noise power spectral density (PSD) estimated from the Gravitational Wave Open Science Center (GWOSC) strain data~\cite{KAGRA:2023pio} using the Welch method implemented in \texttt{PyCBC} with 4\,s segments~\cite{Usman:2015kfa}.final

\subsection{Bayesian comparison with BBHs model }

\begin{figure*}[tbp]
\includegraphics[width=0.95\textwidth]{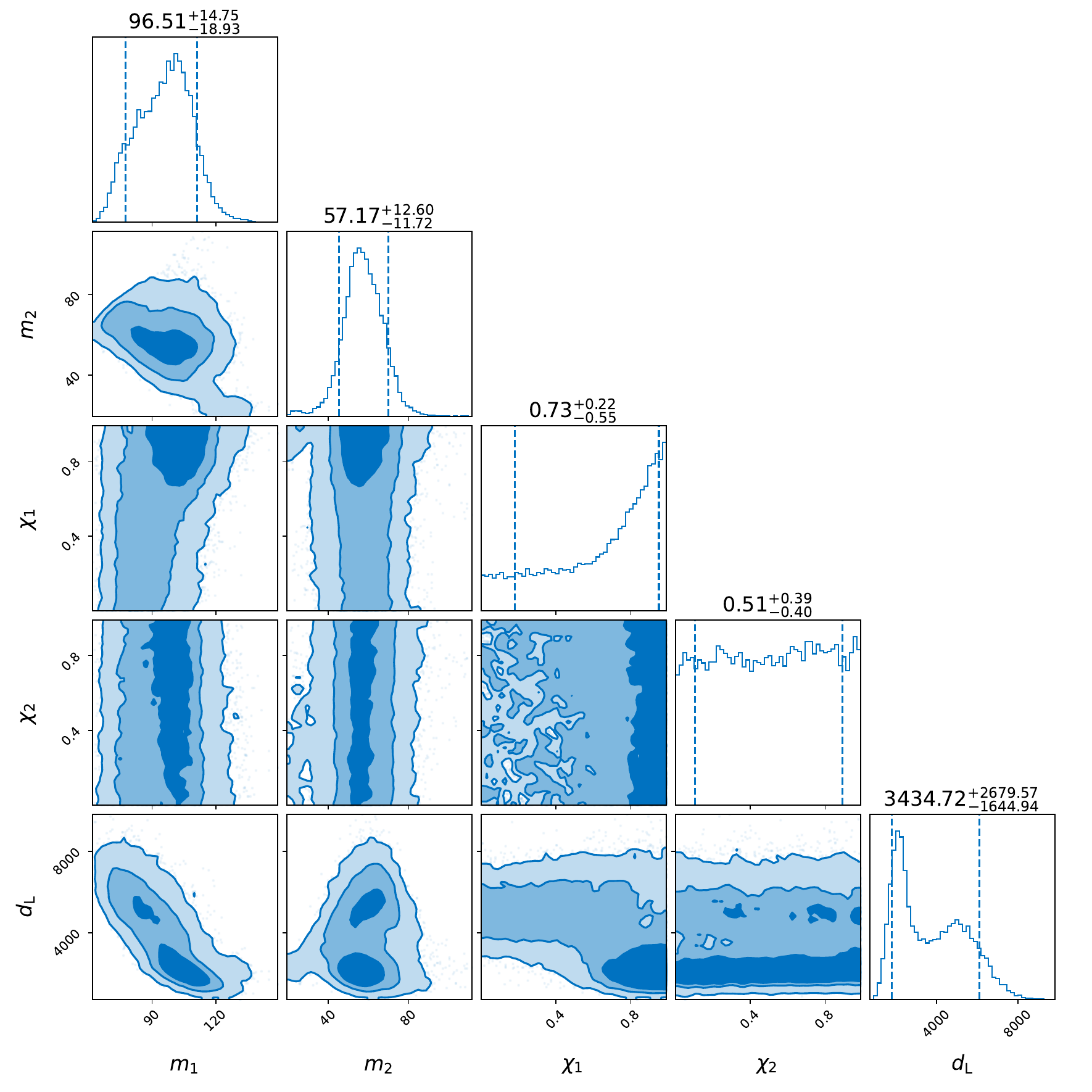}
\caption{The posterior distribution of intrinsic parameters and luminosity distance for the standard
BBHs IMR event with \texttt{IMRPhenomXPHM} waveform.
This distribution is consistent with the result reported by the
LVK collaboration for GW190521 using the same
waveform model~\cite{Estelles:2021jnz}. } \label{IMR_corner}
\end{figure*}

Here, we use the \texttt{IMRPhenomXPHM} as the BBHs waveform
template\footnote{Though the waveform models \texttt{NRSur7dq4}~\cite{Varma:2019csw}
and \texttt{SEOBNRv4PHM}~\cite{Khalil:2023kep,Ramos-Buades:2023ehm,Pompili:2023tna,vandeMeent:2023ols} might yield
more precise parameter estimates, the differences between
different BBHs models are negligible compared to the impact of
distinct source hypotheses on the Bayesian analysis.}, as it
accommodates high mass ratios and spin precession, effectively
covering the parameter space relevant to
GW190521~\cite{Pratten:2020ceb}. In addition, its frequency-domain
construction offers high computational efficiency, facilitating a
systematic comparison with our echo-for-wormhole model. We adopt the same prior distributions as those used in Ref.~\cite{Estelles:2021jnz}, this choice places greater emphasis on unequal–mass systems, thereby improving sampling convergence in the unequal–mass regime. And using
the \texttt{IMRPhenomXPHM} waveform we recover posterior of
relevant parameters\footnote{Here, we also used the localization
reported for ZTF19abanrhr, see Appendix~\ref{App:Sky location}. However, we have verified
that the results are actually nearly same with those without the
ZTF19abanrhr localization. } (see Fig.\ref{IMR_corner}), which consistent with the parameter estimates
reported by the LVK collaboration~\cite{Estelles:2021jnz}, the
merger of BBHs with masses of $m_{1}=97^{+32}_{-21} \mathrm{M}_{\odot}$ and $m_{2}=59^{+22}_{-25} \mathrm{M}_{\odot}$ at a luminosity distance $d_{L}=3.5^{+2.4}_{-2.0}$Gpc.

To evaluate the plausibility of our model as a potential
explanation for the source of GW190521, we compute the SNRs of the
BBHs model and our model for each detector, as well as the network
SNR, by using the publicly available \texttt{PyCBC}
software~\cite{Usman:2015kfa}. The results are summarized in
Table~\ref{SNR_result}. The network SNR is defined as the
quadrature sum of the individual SNRs across the detectors, 
\begin{equation}
    \text{SNR}^{2}_{\text{net}}=\sum_{ifo=\text{H1,L1,V1}}\text{SNR}^{2}_{ifo},
\end{equation}
which serves as a global significance measurement of the event
across the entire detector network, quantifying the combined
detectability of a signal when jointly detected by all
interferometers. Regardless of whether the BBHs model or our model
is applied, the network SNR worked out is significantly above the
noise-only level. The SNRs of our model for each detectors are comparable to those
of the BBHs hypothesis (the network SNR of approximately 14.5 is
also comparable to the SNR of approximately 15.6).

\begin{table}[tbp]
\centering
\begin{tabular}{lcc}
\hline\hline
-- & Echo & BBH \\
\hline
$\rm{SNR}_{\rm{H1}}$  & $7.59$ & $8.57$\\
$\rm{SNR}_{\rm{L1}}$   & $11.84$ & $12.58$\\
$\rm{SNR}_{\rm{V1}}$  & $3.30$&$3.36$\\
$\rm{SNR}_{\rm{net}}$    & $14.45$&$15.59$   \\
\hline\hline
\end{tabular}
\caption{Signal-to-noise ratios for our model and the BBHs model
in each detector (H1, L1, and V1), as well as the corresponding
network SNRs.} \label{SNR_result}
\end{table}

It is also necessary to assess the significance of our
echo-for-wormhole hypothesis relative to the BBHs hypothesis. The
Bayesian model comparison~\cite{Christensen:2022bxb} can be
performed by computing the logarithmic Bayes factor. The log Bayes
factor between the different source hypothesis ($\mathcal{H}_i$)
and the noise-only hypothesis ($\mathcal{H}_0$) is $\mathrm{ln}
\mathcal{B}=\mathrm{ln} {p(d|\mathcal{H}_{i},I)\over
p(d|\mathcal{H}_{0},I)}$,  where $p(d|\mathcal{H}_{i},I)$
represents the evidence for hypothesis $\mathcal{H}_{i}$. By using
the nested sampling package \texttt{dynesty}, we have 
\begin{equation}
    \mathrm{ln} \mathcal{B}^{\text{Echo}}_{\text{BBH}} \simeq -2.9. 
\end{equation}
This negative log Bayes factor indicates that the BBHs hypothesis is favored over our echo-for-wormhole hypothesis by the data. Given the simplified nature of our current echo modeling---namely a nonspinning setup and a first-pulse template without an explicit preceding ringdown---a more complete treatment including rotation and a self-consistent modeling of possible later echoes will be required to fully assess this scenario.

\section{Discussion}
\label{Sec:Conclusion}

In this paper, we investigate the possibility of interpreting
GW190521 as an echo produced by a postmerger wormhole, created
from the inspiral and merger of BBHs in another universe and
connecting it to our own. Our result indicates that our
echo-for-wormhole model yields a network SNR comparable to that of
the BBHs model reported by the LVK collaboration, and our analysis provides a proof-of-principle framework to test such an interpretation for the GW190521 event.

It is usually thought that the ringdown signals of wormhole
consist of a series of continuous echo pulses, here we consider
GW190521 as the first pulse and other undetectable assuming that
the wormhole rapidly pinches off and eventually collapses into a
black hole \cite{Wang:2018mlp} or the amplitudes of subsequent
echoes fall below the sensitivity threshold of current detectors,
e.g.\cite{Abedi:2021tti}. The issues on the physical reliability
of our model are interesting for further studying. The
Morris-Thorne wormhole requires the matter with negative energy
around the wormhole throat \cite{Poisson:1995sv}, which might be
related to current cosmological observations,
e.g.\cite{Ye:2020btb,Wang:2024hwd}. Recent other possible
observations for wormholes have been explored
e.g.\cite{Chen:2022tog,Chen:2024tss}. The current Bayesian results indicate that the BBH hypothesis is favored over our model, and it might be expected that future enhancements to the echo-for-wormhole waveform templates, incorporating more detailed physical characteristics, will be important to assess the robustness of the Bayes factor.

Although the BBH hypothesis is favored for GW190521, how to accurately identify more such short-duration GW events and systematically test alternative hypotheses will be significant. It is noted that recently the
LVK collaboration has reported the detection of
GW231123 on November 23, 2023, an event sharing a similar
burst-like short duration nature with GW190521~\cite{LIGOScientific:2025rsn}, this event has
been motivating intensive investigation on the nature of its
source, e.g.\cite{DeLuca:2025fln,Li:2025fnf,Cuceu:2025fzi,Delfavero:2025lup,Gottlieb:2025ugy,Tanikawa:2025fxw}. Therefore, a systematic model-comparison analysis including
various possible sources for such short duration GW signals might
be necessary for better understanding the physical origins of
corresponding GW events.

\appendix

\section{Sky locations in our analysis}
\label{App:Sky location}

In the initial analysis, we adopted the same sampling configuration as in the main text but did not impose any constraints on the sky location for either model. The parameter estimation results, summarized in Tab.~\ref{tab:skylocation}, show that both models yield sky positions broadly consistent with that of the ZTF19abanrhr flare~\cite{Graham:2020gwr}. To mitigate the impact of source localization uncertainty on Bayesian model selection, we fixed the sky position for both the echo-for-wormhole and BBH models to (RA, DEC)$=(3.35847,~0.60781)$, and computed the Bayes factors before (GW) and after (GW+EM) applying this constraint. The results show that although fixing the sky position increases the Bayes factors for both models, it does not produce a significant change in their relative comparison.
\begin{table}[htb]
  \centering
  \caption{\label{tab:skylocation} 
  Sky locations and Bayes factors for both the BBHs model and our echo-for-wormhole model, compared with the electromagnetic counterpart candidate ZTF19abanrhr. The values for the BBHs model and our model correspond to the maximum-posterior estimates with 90\% credible intervals.}
  \begin{tabular}{lccc}
    \toprule
    Source & Right Ascension [rad] & Declination [rad] & $\ln \mathcal{B}^{\mathrm{GW+EM}}_{\mathrm{GW}}$ \\
    \midrule
    ZTF19abanrhr & 3.35847 & 0.60781 & N/A \\
    BBHs & $3.35^{+2.64}_{-0.14}$ & $0.45^{+0.19}_{-1.34}$ & 3.44 \\
    Echo & $3.52^{+0.60}_{-0.28}$ & $0.62^{+0.28}_{-1.38}$ & 3.21 \\
    \bottomrule
  \end{tabular}
\end{table}

\acknowledgments

We thank Jun Zhang for the comment on our manuscript, and also
Jahed Abedi, Niayesh Afshordi for valuable discussions on issues
relevant to the GW echoes. This work is supported by National Key
Research and Development Program of China, No.2021YFC2203004, and
NSFC, No.12475064, and the Fundamental Research Funds for the
Central Universities.
%


 \end{document}